\documentclass[useAMS,usenatbib]{mn2e}
\usepackage{natbib}
\usepackage{lineno,hyperref}
\modulolinenumbers[5]

\usepackage{amsmath}
\usepackage{amsfonts}
\usepackage{amssymb}

\usepackage{graphicx}
\usepackage{epsfig}
\usepackage{verbatim}

\def\m#1{\mbox{$#1$}} %My new command for equations in the text.

\title[]{Extragalactic magnetic fields unlikely generated at the electroweak phase transition}

\author[]{Jacques M. Wagstaff\thanks{jwagstaff@hs.uni-hamburg.de} and Robi Banerjee\\
Hamburger Sternwarte, Gojenbergsweg 112, 21029 Hamburg, Germany}

\begin{document}
\date{}

\pagerange{\pageref{firstpage}--\pageref{lastpage}} \pubyear{2015}

\maketitle

\label{firstpage}

\begin{abstract}
In this letter we show that magnetic fields generated at the electroweak phase transition are most likely too weak to explain the void magnetic fields apparently observed today unless they have considerable helicity. We show that, in the simplest estimates, the helicity naturally produced in conjunction with the baryon asymmetry is too small to explain observations, which require a helicity fraction at least of order $10^{-14}-10^{-10}$ depending on the void fields constraint used. Therefore new mechanisms to generate primordial helicity are required if magnetic fields generated during the electroweak phase transition should explain the extragalactic fields. 
\end{abstract}


%%---------------------------------------------------------
%%---------------------------------------------------------

\section{Introduction}
%\paragraph{Introduction:}

The evolution of magnetic fields depends strongly on its helicity. 
On the one hand, the global conservation of magnetic helicity directly leads to an inverse cascade of energy from small scales to large scales e.g. \citep{Biskamp1993,Christensson:2000sp,Banerjee:2004df}. On the other hand, the helicity conservation leads to slower decay of the magnetic energy compared to the non-helical case, potentially producing stronger fields at present day. This combination of effects could prove to be of great importance for the explanation of large scale magnetic fields observed in the Universe today \citep{Banerjee:2004df,Tashiro:2013ita}.
In this short note we aim to constrain the primordial magnetic helicity from the apparent observations of void magnetic fields. We show that magnetic fields generated at the electroweak phase transition (EWPT) must have considerable helicity in order to explain the extragalactic magnetic fields and that the helicity density arising naturally with electroweak baryogenesis is too weak to be the dominant source of magnetic helicity (see \citep{Sigl:2002kt} for an early discussion on this topic). 
Our conclusions are based on well established results from magnetohydrodynamic (MHD) turbulence for the magnetic field decay rates \citep{Banerjee:2004df,Campanelli:2007tc,Campanelli:2013iaa} and the simplest estimates of magnetic helicity generation by \citep{Vachaspati:2001nb}. Any deviations from these conclusions would require deviations from such simple assumptions 
(see e.g. \citep{Kahniashvili:2012uj,Brandenburg:2014mwa}), which we discuss later in this work.

\section{Basic equations}
%\paragraph{Basic equations:}

We start by writing down the two-point correlation function for a statistically homogeneous and isotropic stochastic magnetic field (see e.g. \citep{Durrer:2013pga} and references therein)
\begin{align}
 \langle B_i(\bmath{k})B_j^*&(\bmath{k}')\rangle=
  (2\pi)^3\delta(\bmath{k}-\bmath{k}')\times\nonumber\\
  &\times\left[\left(\delta_{ij}-\hat k_i\hat k_j\right)P_B(k)
  +i\epsilon_{ijk}\hat k_k P_H(k)\right]\,.
\end{align}
The above spectrum has been decomposed into parity conserving and parity violating components, or \emph{magnetic} and \emph{helical} spectra. The magnetic spectrum is given by
\m{P_B(k)=\langle |\bmath{B}(\bmath{k})|^2\rangle/2\equiv(\rho/k^2)\langle M_k\rangle},
%
\begin{comment}
\begin{equation}
 P_B(k)
  =\frac{1}{2}\langle |\bmath{B}(\bmath{k})|^2\rangle
  \equiv\frac{\rho}{k^2}\langle M_k\rangle\,,
\end{equation}
\end{comment}
%
which depends only on the amplitude of $\bmath k$ not its direction. We  define the spectrum $M_k$ above to match the conventions of \citep{Saveliev:2012ea,Saveliev:2013uva}, where $\rho$ is the total energy density and comoving quantities are used throughout.
Then, assuming a power law \m{P_B(k)\simeq P_B^0k^n} on large scales, we can volume average on a region of size $L^3$ to estimate the average field on a given scale $L$~\citep{CausalBfield}
\begin{equation}\label{eq:BL_largeSacles}
 B^2_L\equiv\langle\bmath{B}^2_L(\bmath{x})\rangle=
  \frac{P_B^0}{2\pi^2}\frac{1}{L^{n+3}}\Gamma\left(\frac{n+3}{2}\right)\,.
\end{equation}
Causality restricts the power law index $n$ to be an even integer \m{n\geq2}~\citep{CausalBfield}. For the limiting case \m{n=2}, which is the expected scaling, one finds that \m{B_L\propto L^{-5/2}}.
In \citep{Saveliev:2012ea} the authors showed numerically that, independent of the turbulent flow, a large scale magnetic field tail develops with the scaling \m{B_L\sim L^{-5/2}}, in agreement with the causality constrained averaged field described above.

\begin{comment}
The two-point correlation function can be written as
\begin{equation}
 \langle B_i(\bmath{k})B_j^*(\bmath{k}')\rangle=
  \frac{(2\pi)^3}{V}\delta(\bmath{k}-\bmath{k}')\frac{\rho}{k^2}
  \left[\left(\delta_{ij}-\hat k_i\hat k_j\right)\langle M_k\rangle
  -\frac{ik}{8\pi}\epsilon_{ijk}\hat k_k \langle \mathcal H_k\rangle\right]\,,
\end{equation}
\end{comment}

The averaged magnetic energy density is obtained by integrating over the \emph{local} energy density \m{u_B=\bmath{B}^2/8\pi}, i.e. 
\begin{equation}\label{eq:epsilon_B}
 \epsilon_B=\frac{1}{V}\int u_B\mathrm{d}\bmath{r}
  %=\frac{1}{8\pi V}\int\bmath{B}^2(\bmath{r})\mathrm{d}\bmath{r}
  =\frac{1}{8\pi}\int|\bmath{B}(\bmath{k})|^2\mathrm{d}\bmath{k}
  = \rho\int M_k\mathrm{d}k
  \,.
\end{equation}
Assuming that the magnetic energy is concentrated at the \emph{integral} scale (denoted by the index `$I$'), which is the peak of the spectrum in Fourier space, we can write
\m{\epsilon_B=\rho\int k M_k\mathrm{d}\ln k\simeq\rho k_I M_I}.
%
\begin{comment}
\begin{equation}\label{eq:epsilon_B2}
 \epsilon_B=\rho\int k M_k\mathrm{d}\ln k\simeq\rho k_I M_I
  \,.
\end{equation}
\end{comment}
%
%In Eqs.~(\ref{eq:epsilon_B}) and (\ref{eq:epsilon_B2})
In the above we adopt the conventions from \citep{Saveliev:2012ea}, where the authors showed numerically that indeed most of the energy is concentrated at the integral scale. We can also define an effective magnetic field strength 
\m{\epsilon_B=\int\mathrm{d}\ln k(B_k^{\mathrm{eff}})^2/8\pi}.
%
\begin{comment}
\begin{equation}
 \epsilon_B=\rho\int k M_k\mathrm{d}\ln k
  \equiv \int\frac{(B_k^{\mathrm{eff}})^2}{8\pi}\mathrm{d}\ln k\,.
\end{equation}
\end{comment}
%
In the above we set the wavenumber $k=2\pi/L$ corresponding to the scale $L$ in Eq.~(\ref{eq:BL_largeSacles}). We can also identify the integral scale as the coherence length ($\lambda_B$), 
and the \emph{effective} magnetic field as the observed magnetic field strength ($B_\lambda$), hence
\m{M_I=B^2_I/8\pi\rho k_I=B^2_\lambda\lambda_B/16\pi^2\rho}.
%
\begin{comment}
\begin{equation}
 M_I=\frac{B^2_I}{8\pi\rho k_I}=\frac{B^2_\lambda\lambda_B}{16\pi^2\rho}\,.
\end{equation}
\end{comment}
%

The helical part of the spectrum is determined by
\begin{equation}
  P_H(k)=-\frac{i}{2}\langle(\bmath{\hat k}\times\bmath{B}(\bmath{k}))
  \cdot\bmath{B}^*(\bmath{k}) \rangle
  \equiv-\frac{\rho}{8\pi k}\langle \mathcal H_k\rangle\,,
\end{equation}
where again the helical spectrum $\mathcal H_k$ is defined following the conventions of \citep{Saveliev:2013uva}. On any given scale $k$ there is a \emph{realizability} condition given by \m{|\mathcal H_k|\leq 8\pi M_k/k},
%
\begin{comment}
\begin{equation}
 |\mathcal H_k|\leq \frac{8\pi}{k} M_k
 \,,
\end{equation}
\end{comment}
%
from which we can define \m{f\equiv k\mathcal H_k/8\pi M_k} as the helicity fraction,
where $f=0$ for the non-helical case and $f=1$ for the maximally helical case.
The average helicity density is given by
\begin{equation}
 h_B=\frac{1}{V}\int\left(\bmath{A}\cdot\bmath{B}\right)\mathrm{d}\bmath{r}
  =\rho\int\mathcal{H}_k\mathrm{d}k\simeq\rho k_I\mathcal{H}_I
  \,,
\end{equation}
where \m{\bmath{B}=\nabla\times\bmath{A}}, and in the last equality we also assume that the helicity density is concentrated at the integral scale. Magnetic helicity is a useful quantity since it is conserved \m{h_B\simeq\mathrm{const.}} in the early Universe when the conductivity is very large~\citep{Biskamp1993}.

\section{Basic constraints}
%\paragraph{Basic constraints:}

Let us first consider the constraints on void magnetic fields from the $\gamma$-ray observations of TeV Blazars~\citep{Neronov:1900zz,Taylor:2011bn,Tavecchio:2010mk,Dolag:2010ni,Essey:IGMF2010,Dermer_IGMF2011}. Authors in \citep{Taylor:2011bn} showed that the minimum magnetic field strength depends on the mechanism of suppression of the cascade signal. For suppression due to time delay, the minimum required field strength is \m{\sim10^{-17}}~G, whereas for the extended emission they find \m{\sim10^{-15}}~G (see also \citep{Tavecchio:2010mk,Essey:IGMF2010,Dermer_IGMF2011}). The above bounds become tighter as \m{\lambda_B^{-1/2}} for scales smaller than \m{\sim1}~Mpc.
We note that the above observations are not conclusive~\citep{Arlen:2012iy,Broderick_etal_2012}, and the authors in \citep{Taylor:2011bn} stress that, in any case, the bounds should be taken as an order-of-magnitude estimate. However, in this work we assume that the above constraints are actual bounds on void magnetic fields. For the purpose of this work we shall consider the bounds~\citep{Taylor:2011bn,Dermer_IGMF2011}
%
%
\begin{comment}
\begin{equation}\label{eq:Fermi}
 B_\lambda\gtrsim \left\{  
\begin{array}{l l}
10^{-15}~\textrm{G}\,, & \lambda_B\gtrsim0.1~\textrm{Mpc}\\
10^{-15}\lambda_B^{-\frac12}~\textrm{G}\,, & \lambda_B\lesssim0.1~\textrm{Mpc}\,,
 \end{array}
\right.
\end{equation}
\end{comment}
%
\begin{equation}\label{eq:Fermi}
 B_\lambda\gtrsim 
(10^{-15}-10^{-18})~\lambda_B^{-\frac12}~\textrm{G}\,, 
\quad\lambda_B\lesssim1~\textrm{Mpc}\,,
\end{equation}
and later comment on our conclusions if this bound is relaxed somewhat.

\begin{comment}
\begin{equation}\label{eq:Fermi}
 B_\lambda\gtrsim \left\{  
\begin{array}{l l}
3\times10^{-16}~\textrm{G}\,, & \lambda_B\gtrsim0.1~\textrm{Mpc}\\
3\times10^{-16}\lambda_B^{-\frac12}~\textrm{G}\,, & \lambda_B\lesssim0.1~\textrm{Mpc}\,.
  \end{array}
\right.
\end{equation}
\end{comment}

The second constraint comes from energy considerations. The initial magnetic energy 
at magnetogenesis (denoted by the index `$*$') can at most be in equipartition 
with radiation, i.e. \m{u_B=\rho/2}, hence
\begin{equation}
 B_{\lambda,*}\leq B_{\lambda,*}^{\mathrm{max}}\equiv (4\pi\rho)^{\frac12}
  \simeq 3\times10^{-6}~\textrm{G}\,,
\end{equation}
where the radiation here is taken to be the CMB photons~\citep{Banerjee:2004df}. We note that this bound satisfies the constraint from Big Bang Nucleosynthesis~\citep{Kahniashvili:2009qi}.

For magnetic fields generated at a time during the radiation dominated era (in contrast to inflationary magnetogenesis), the basic constraint on the coherence length is the horizon size at the time of magnetogenesis
\begin{equation}
 \lambda_{B,*}\leq \lambda^{\mathrm{max}}_{B,*}\equiv \frac{1}{aH}\Big|_*\,.
\end{equation}
The horizon size is \m{2\times10^{-10}}~Mpc and \m{3\times10^{-7}}~Mpc at the electroweak and QCD  phase transitions respectively. 

For the purpose of this paper, these are the only constraints that we need to consider.

%%-------------------------------------------------------------------------
\section{Magnetic field evolution}
%\paragraph{Magnetic field evolution:}

The magnetic field strength and coherence length evolve during the radiation dominated era due to turbulent MHD effects \citep{Sigl:2002kt,Banerjee:2004df,Campanelli_free-turb,Durrer:2013pga,Wagstaff:2013yna}. Such effects include the free turbulent decay of magnetic fields, which is what we are mostly concerned with here, but we note that the turbulent amplification of weak fields by the small-scale dynamo is also possible~\citep{Wagstaff:2013yna}.
In this paper we quote well known results from MHD turbulence and in particular we use the decay laws from the detailed studies in \citep{Banerjee:2004df,Campanelli:2007tc,Campanelli:2013iaa,Saveliev:2012ea,Saveliev:2013uva}.

If the initial spectral helicity is negligible \m{f_*\ll1}, 
there is a \emph{direct cascade} of energy due to the selective decay of modes in $k$-space. Indeed, the large $k$-modes are dissipated and decay so that the integral scale evolves down along the large scale spectrum. The general decay law for the magnetic energy is \m{M_I\propto t^{-2(n'-1)/(n'+2)}} and for the integral scale \m{L_I\propto t^{2/(n'+2)}}, where \m{n'\equiv n+3}. These decay laws are obtained through analytical considerations in \citep{Banerjee:2004df,Saveliev:2012ea,Saveliev:2013uva,Campanelli:2007tc,Campanelli:2013iaa} and confirmed numerically in \citep{Banerjee:2004df}. 
From this it can be shown that \m{k_I\propto a^{-2/7}}, \m{M_I\propto a^{-8/7}}, and 
\m{\mathcal H_I\propto a^{2/7}}, where $a$ is the scale factor and \m{n=2} is used in Eq.~(\ref{eq:BL_largeSacles}) due to causality constraints for the large scale spectrum~\citep{Banerjee:2004df,Saveliev:2012ea,Saveliev:2013uva,CausalBfield}. Hence, the magnetic field strength on the integral scale evolves as
\begin{equation}\label{eq:BL_nonHelical}
 B_I\propto \lambda_I^{-\frac52}\,.
\end{equation} 
Therefore, the above scaling due to the evolution decay law on the integral scale coincides with the scaling of the smoothed magnetic field in Eq.~(\ref{eq:BL_largeSacles}). The smoothed magnetic field $B_L$ on a scale $L$ is equivalent to the magnetic field on the integral scale $B_I$, assuming that most of the magnetic energy resides on the integral scale. The point is that, as the magnetic energy on small scales dissipate, the integral scale field $B_I$ moves down along the large scale spectrum, hence the above scaling.

Here we comment on the very interesting and exciting new developments in turbulent MHD where an apparent “inverse transfer” of magnetic energy occurs for the non-helical case. Due to this effect the authors in \citep{Kahniashvili:2012uj} obtain a weaker evolution for non-helical magnetic fields \m{B_I\propto t^{-1/2}} and \m{L_I\propto t^{1/2}} giving the relation \m{B_I\propto\lambda_I^{-1}} [c.f Eq.~(\ref{eq:BL_nonHelical})] (see also \citep{Campanelli:2004wm}). This numerically observed effect has also been studied in \citep{Brandenburg:2014mwa} with the conditions of high resolutions, and magnetically dominant turbulence. However the condition of magnetically dominant turbulence is perhaps not satisfied in the early Universe. Magnetogenesis at first order phase transitions typically produce a lot of turbulent kinetic energy. The generated magnetic field, through dynamo action, comes into equipartition with the kinetic energy, but is unlikely to dominate over the kinetic energy e.g. \citep{B-phaseT}. Furthermore, in the study by \citep{Brandenburg:2014mwa} it seems that the inverse transfer is less efficient for large Prandtl numbers, but the Prandtl numbers in the early Universe are huge.
We clearly state here that for the following arguments we assume the decay laws of \citep{Banerjee:2004df,Campanelli:2007tc,Campanelli:2013iaa} in Eq.~(\ref{eq:BL_nonHelical}), affirming that our assumptions could be challenged due to the important works of \citep{Kahniashvili:2012uj,Brandenburg:2014mwa}.

If the helicity density is non-zero, the helicity fraction in the regime \m{f\ll1} evolves as \m{f=k_I\mathcal{H}_I/8\pi M_I\propto a^{8/7}}. This evolution occurs until the time of recombination, or
until a state of maximum helicity is reached \m{f=1}, whichever comes first~\citep{Banerjee:2004df,Saveliev:2013uva}. The evolution of the field strength and coherence length essentially ceases, i.e. with logarithmic scaling, when the Universe becomes matter dominated~\citep{Banerjee:2004df}. Following literature we approximate this time as the time of recombination.

When the maximal helicity \m{f=1} is reached, an \emph{inverse cascade} occurs where magnetic energy is transferred from small scales to large scales \citep{Christensson:2000sp}. When applying helicity conservation \m{h_B\simeq\mathrm{const.}}, remarkably the decay law becomes independent of the large scale slope and it can be shown that 
\m{k_I\propto a^{-2/3}}, \m{M_I\simeq \textrm{const.}}, \m{\mathcal H_I\propto a^{2/3}}, and
the fraction remains constant \m{f=1}~\citep{Christensson:2000sp,Banerjee:2004df,Saveliev:2013uva}. Hence, the field strength at the integral scale evolves as 
\begin{equation}
 B_I\propto \lambda_I^{-\frac12}\,,
\end{equation}
which by coincidence is the same scaling as the Fermi observational bound in Eq.~(\ref{eq:Fermi}). This `weaker' evolution of a maximally helical field in the radiation era can have important consequences for magnetic fields observed today.

\begin{figure}
  \includegraphics[width=90mm]{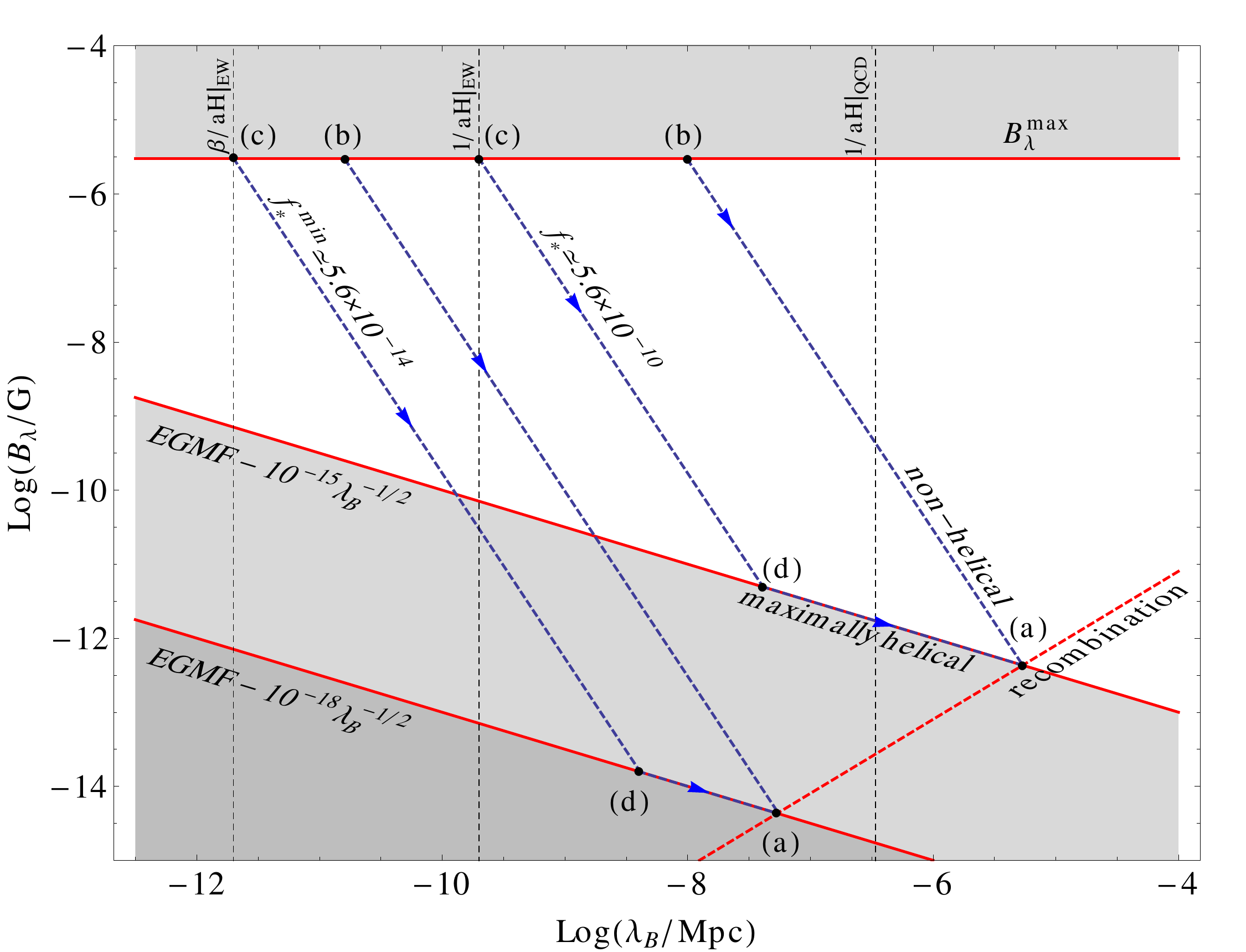}
\caption{\label{fig:ConstraintsPlot} 
In the greyed out regions, constraints on present day magnetic fields are shown from Fermi observations of $\gamma$-ray sources (see text above Eq.~(\ref{eq:Fermi}) for the different constraints) and an upper bound set from energy considerations. Fields generated in the radiation era evolve to the line labelled ``recombination''. The evolutionary tracks from magnetogenesis until recombination are marked by dashed lines and depend on the helicity fraction $f_*$. The minimum field configuration at recombination to explain the void fields is marked by point (a). If there is zero helicity, the field configuration at megnetogenesis is marked by point (b). With non-zero helicity the initial field configuration can be reduced e.g point (c), where the field becomes maximally helical at point (d). 
}
\end{figure}

We summarise the above results in Fig.~\ref{fig:ConstraintsPlot} where the magnetic field evolves until recombination (index `$\mathrm{rec}$'), and the final field configuration falls on the line \citep{Banerjee:2004df,Durrer:2013pga} (labelled `recombination' in Fig.~\ref{fig:ConstraintsPlot})
\m{B_{\lambda,\mathrm{rec}}/\mathrm{G}\simeq8\times10^{-8}\lambda_{B,\mathrm{rec}}/\mathrm{Mpc}}.
This line corresponds to the largest eddies being processed at recombination 
\m{1/(aH)|_{\mathrm{rec}}\simeq\lambda/v_A} with $v_A$ the Alfv\'{e}n speed \citep{Jedamzik:1996wp,Banerjee:2004df}.
Hence, we obtain our first constraint. 
Magnetic fields generated during the radiation era will evolve to fall on the above line at recombination, which are also the values that will be observed today since the field strength and coherence length do not evolve significantly in the matter dominated Universe. In order for such fields to explain the Fermi observations it is required that
\m{B_{\lambda,\mathrm{rec}}\geq B_{\lambda,\mathrm{rec}}^{\mathrm{min}}= 4.3\times10^{-13}~\mathrm{G}} and 
\m{\lambda_{B,\mathrm{rec}}\geq\lambda_{B,\mathrm{rec}}^{\mathrm{min}}=   5.4\times10^{-6}~\mathrm{Mpc}} (using the stronger bound in Eq.~(\ref{eq:Fermi})).
%
\begin{comment}
\begin{eqnarray}
 B_{\lambda,\mathrm{rec}} &\geq& B_{\lambda,\mathrm{rec}}^{\mathrm{min}}= 
  2.0\times10^{-13}~\mathrm{G}\,,\nonumber\\
  %\qquad\textrm{and}\qquad
  \lambda_{B,\mathrm{rec}} &\geq &\lambda_{B,\mathrm{rec}}^{\mathrm{min}}= 
  2.5\times10^{-6}~\mathrm{Mpc}\,.
\end{eqnarray}
\end{comment}
%
This minimum field configuration is labelled by point (a) in Fig.~\ref{fig:ConstraintsPlot}. From this minimum configuration we trace the evolution back to the time of magnetogenesis 
and find the minimum values for the field strength and coherence length at those times. 
We find that, if the field has zero initial helicity \m{f_*=0}, the initial field 
configuration must be (point (b) in Fig.~\ref{fig:ConstraintsPlot})
\m{B_{\lambda,*}\leq B_{\lambda,*}^{\mathrm{max}}= 3\times10^{-6}~\mathrm{G}} and
\m{\lambda_{B,*}\geq 9.9\times10^{-9}~\mathrm{Mpc}}.
%
\begin{comment}
\begin{eqnarray}
 B_{\lambda,*}&\leq& B_{\lambda,*}^{\mathrm{max}}= 3\times10^{-6}~\mathrm{G}\,,\nonumber\\%\quad
%\qquad\textrm{and}\qquad
\lambda_{B,*}&\geq& 3.4\times10^{-9}~\mathrm{Mpc}\,.
\end{eqnarray}
\end{comment}
%
This coherence length is smaller than the horizon size at the QCDPT
\m{\sim10^{-7}~\mathrm{Mpc}}, but larger than the horizon size at the EWPT
\m{\sim10^{-10}~\mathrm{Mpc}}. Hence, we come to our first important conclusion. With the stronger bound in Eq.~(\ref{eq:Fermi}), it is impossible to generate magnetic fields at the EWPT which can explain the
apparently observed void magnetic fields if the magnetic fields have zero helicity.

To make this point even stronger we can ask the question: how far must the bound from $\gamma$-ray observations go down so that magnetic fields with zero helicity generated at the EWPT produce the void fields? Magnetic fields generated at the EWPT are constrained by (labelled by point (c) in Fig.~\ref{fig:ConstraintsPlot})
\m{B_{\lambda,*}\leq B_{\lambda,*}^{\mathrm{max}}= 3\times10^{-6}~\mathrm{G}} and 
\m{\lambda_{B,*}\leq\lambda_{\mathrm{EW}}=2\times10^{-10}~\mathrm{Mpc}}.
%
\begin{comment}
\begin{eqnarray}
 B_{\lambda,*}&\leq& B_{\lambda,*}^{\mathrm{max}}= 3\times10^{-6}~\mathrm{G}\,,\nonumber\\
  \lambda_{B,*}&\leq& 2\times10^{-10}~\mathrm{Mpc}\,.
\end{eqnarray}
\end{comment}
%
With this we find that the new hypothetical ``Fermi'' constraint should be
\begin{equation}
 B_\lambda\gtrsim1.5\times10^{-17}
  \left(\frac{B_{\lambda,*}}{B_{\lambda,*}^{\mathrm{max}}}\right)^{\frac37}
  \left(\frac{\lambda_{B,*}}{\lambda_{\mathrm{EW}}}\right)^{\frac{15}{14}}
  \lambda_B^{-\frac12}~\mathrm{G}\,.
\end{equation}
This bound is compatible with the weakest constraint from $\gamma$-ray observations \m{B_\lambda\gtrsim10^{-17}} due to time delay suppression~\citep{Taylor:2011bn} and the bound found in \citep{Dermer_IGMF2011} \m{B_\lambda\gtrsim10^{-18}}. 
However, the expected coherence length of magnetic fields generated at the EWPT is roughly of order the bubble size, which is somewhat smaller than the horizon size by a factor \m{\beta\equiv\lambda_{B,*}/\lambda_{\mathrm{EW}}\sim10^{-2}}~\citep{Turok:1992jp}. And, since $B_{\lambda,*}$ is expected to be a few orders of magnitude below $B_{\lambda,*}^{\mathrm{max}}$~\citep{Baym:1995fk}, we can see that the existing Fermi bound would have to decrease considerably. Hence, even with the weakest constraints on void magnetic fields, fields generated at the EWPT most probably require helicity in order to explain the observed void fields.

%\begin{comment}
%\begin{widetext}
%\end{widetext}
%\end{comment}

Let us assume that magnetic fields are generated at the EWPT~\citep{Baym:1995fk}. If the initial helicity density is non-zero and the helicity fraction is large enough, then the inverse cascade may take effect and make it possible to explain the void fields. As seen above, when the spectral helicity is small, the fraction evolves as 
\m{f\propto a^{8/7}} until a state of maximum helicity is reached 
\m{f_{\mathrm{tr}}=1} at the \emph{transition} time, therefore
\m{f_*=(a_*/a_{\mathrm{tr}})^{8/7}=(\lambda_{I,*}/\lambda_{I,\mathrm{tr}})^4}.
%
\begin{comment}
\begin{equation}
 f_*=\left(\frac{a_*}{a_{\mathrm{tr}}}\right)^{\frac87}
  =\left(\frac{\lambda_{I,*}}{\lambda_{I,\mathrm{tr}}}\right)^{4}\,.
\end{equation}
\end{comment}
%
To find the minimum helicity fraction required in order to explain the Fermi observations, we note that $\lambda_{I,\mathrm{tr}}$ falls on the Fermi constrain in Eq.~(\ref{eq:Fermi}) (labelled by point (d) in Fig.~\ref{fig:ConstraintsPlot}) since the Fermi constraint and the evolution for a maximally helical field has the same scaling $\lambda_B^{-1/2}$. 
From the above considerations we find that
\begin{equation}\label{eq:fmin}
 f_*\geq f^{\mathrm{min}}_{*}=
  \left(5.6\times10^{-10}\right)\frac{\lambda_\mathrm{EW}}{\lambda_{I,*}}
  \left(\frac{B^{\mathrm{max}}_{\lambda,*}}{B_{I,*}}\right)^{2}
  %=\left(1\times10^{-11}\right)\frac{\lambda_\mathrm{EW}}{\lambda_{I,*}}
  %\frac{1}{2f_E}
\,,
\end{equation}
in order to explain the void magnetic fields. 
The above constraint is obtained using the stronger bound in Eq.~(\ref{eq:Fermi}), with the weaker bound and \m{\beta=10^{-2}}, the helicity fraction reduces to \m{f^{\mathrm{min}}_{*}\simeq5.6\times10^{-14}}.
We can also constrain the average helicity density, which is given by
\m{h_B\simeq\rho k_I\mathcal{H}_I=8\pi\rho fM_I=fB_I^2/k_I}.
%
\begin{comment}
\begin{equation}
 h_B\simeq\rho k_I\mathcal{H}_I
  =8\pi\rho fM_I=f\frac{B_I^2}{k_I}\,.
  %=\frac{(B_{\lambda,*}^{\mathrm{max}})^2}{4\pi} k_I\mathcal{H}_I\,.
\end{equation}
\end{comment}
%
Since the helicity density is a conserved quantity \m{h_B\simeq\textrm{const.}}, we find that
\m{h_{B,*}\simeq h_{B,\mathrm{rec}}}.
%
\begin{comment}
\begin{equation}
 h_{B}\simeq h_{B,*}\simeq f_*\frac{\lambda_{I,*}}{2\pi}B^2_{I,*}
  \simeq
  h_{B,\mathrm{rec}}\simeq \frac{\lambda_{I,\mathrm{rec}}}{2\pi}B^2_{I,\mathrm{rec}}\,.
\end{equation}
\end{comment}
%
Hence, the minimum helicity density required to explain the Fermi observations is
\begin{equation}
 h^{\mathrm{min}}_{B}\simeq f^{\mathrm{min}}_*\frac{\lambda_{I,*}}{2\pi}B^2_{I,*}
 % =\left(1\times10^{-11}\right)\frac{\lambda_\mathrm{EW}}{2\pi}
 % \left(B^{\mathrm{max}}_{\lambda,*}\right)^{2}
  \simeq 1.6\times10^{-13}~\mathrm{nG}^2\mathrm{Mpc}
  \,,
\end{equation}
which goes down to \m{h^{\mathrm{min}}_{B}\simeq 1.6\times10^{-19}~\mathrm{nG}^2\mathrm{Mpc}}
when considering the weaker bound in Eq.~(\ref{eq:Fermi}) and \m{\beta=10^{-2}}.

In \citep{Vachaspati:2001nb} the author estimated the primordial magnetic field helicity generated at electroweak baryogenesis. The production of baryon number requires changes in the Chern-Simons number, which are generated by the production and dissipation of nonperturbative field configurations, e.g. linked loops of electroweak strings. Such configurations would decay in the true vacuum phase of the EW transition leaving behind linked magnetic field lines. Hence the connection between baryon number and magnetic helicity. Indeed, the change in magnetic helicity is \m{\sim10^2} for every baryon produced~\citep{Vachaspati:2001nb}. Hence, the helicity density can be estimated today as \m{h_B\sim 10^2 n_b}, where \m{n_b\sim10^{-6}~\mathrm{cm}^{-3}} is the baryon density observed today, therefore the helicity density is estimated as \m{h_B\sim10^{-27}~\mathrm{nG}^2\mathrm{Mpc}}, i.e. \m{f_*\sim10^{-24}} assuming \m{B^{\mathrm{max}}_{\lambda,*}} and \m{\lambda_\mathrm{EW}}. The length scale in which helicity is expected to be maximal can be estimated by considering the 
characteristic length scale of the 
gauge field configurations \m{L\sim1/e^2T_{\mathrm{EW}}} \citep{Vachaspati:2001nb} which is much smaller than the horizon size at the EW scale by a factor \m{\sim10^{-17}}. Therefore, with no other sources for generating magnetic helicity other than this simple mechanism from electroweak baryogenesis, we show that magnetic fields generated at the EWPT most probably cannot explain the void magnetic fields observed today. However, there are some exciting new ideas regarding Chiral MHD which can excitep helical magnetic fields at very hight temperatures \citep{Boyarsky:2011uy,Boyarsky:2015faa}. It will be interesting to see how such mechanisms can affect our conclusions, this will be investigated in future publications.

%%-------------------------------------------------------------------------------------
\section{Constraints from first-order phase transitions}
%\paragraph{Constraints from first-order phase transitions:}

The EWPT could be a first-order transition in certain extensions to the Standard Model (see e.g. \citep{Laine:1998qk}). Such models can therefore be constrained by extragalactic magnetic fields, since their parameters, which characterise the phase transition, also determine the minimum helicity fraction required to produce the void fields.
%
%
\begin{comment}
The minimum helicity fraction clearly depends on the energy budget of the phase transition that can go into generating magnetic fields. In the above analysis we assume initial equipartition of magnetic and kinetic energy \m{u_B\simeq u_K}. The magnetic energy is some
fraction $f_E$ of the radiation energy \m{f_E\equiv u_B/\rho}, which has a maximum value \m{f_E=1/2} for initial equipartition, hence \m{f_E=u_K/\rho}.
\end{comment}
%
%

\begin{figure}
  \includegraphics[width=90mm]{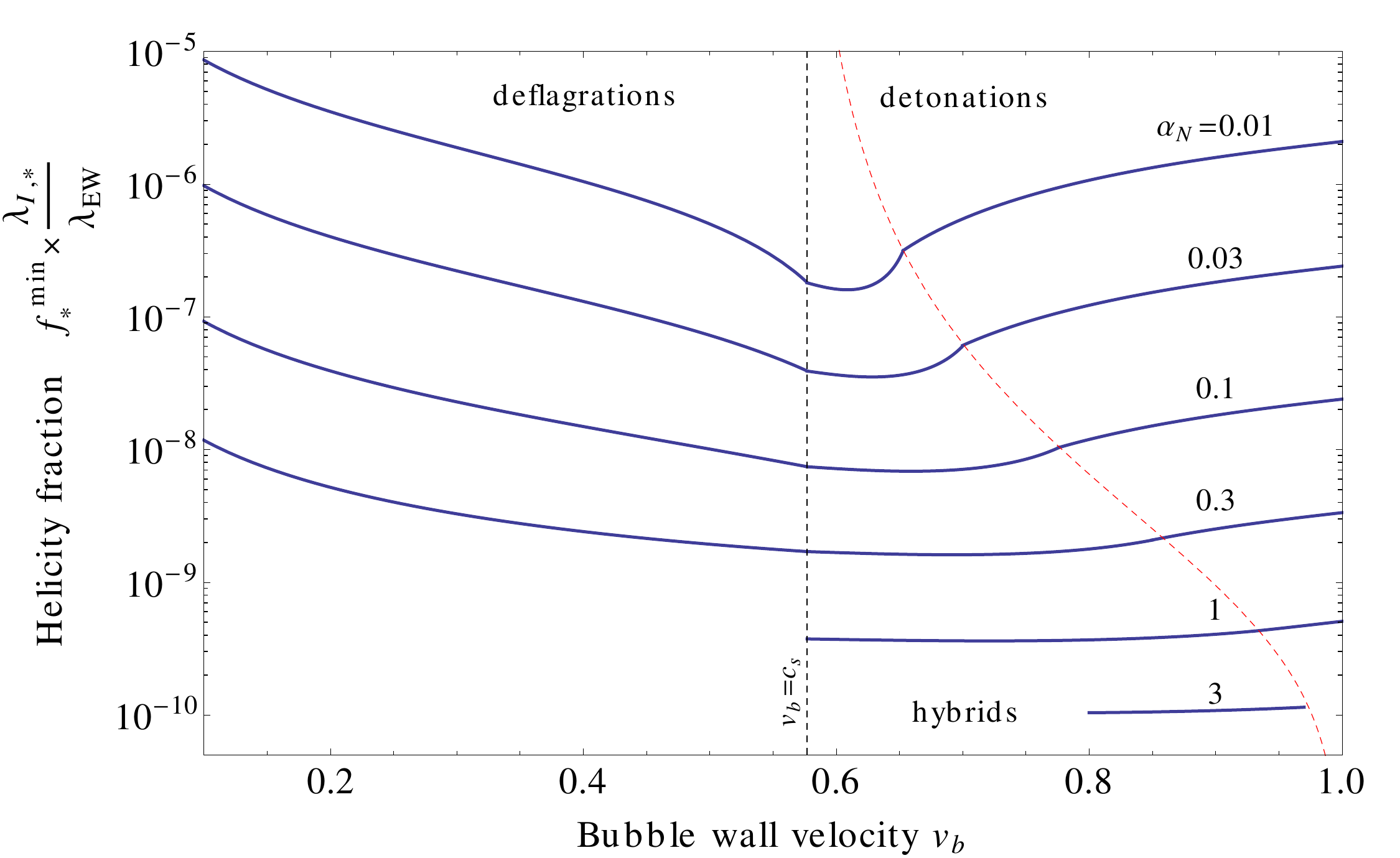}
\caption{
\label{fig:PT_Plot}
Depending on the phase transition parameters, i.e. the strength of the transition $\alpha_N$ and the bubble wall velocity $v_b$, the efficiency parameter $\kappa$ can be determined which in turn determines the energy of the phase transition that goes into producing magnetic fields. From this we can determine the minimum helicity fraction $f^{\mathrm{min}}_{*}$ required to produce the void magnetic fields given a set of model parameters. 
%This work takes into account the analytical and numerical work on first-order phase %transitions in Ref.~\citep{Espinosa:2010hh}.
}
\end{figure}

Three parameters characterise model-independent analysis of first-order phase transitions~\citep{Steinhardt1982,Espinosa:2010hh}. 
The first parameter \m{\alpha_N\equiv\epsilon_{\mathrm{vac}}/\rho_{\mathrm{rad}}} is the ratio of the vacuum energy to the radiation energy density, which characterises the strength of the phase transition. The second is the efficiency parameter \m{\kappa\equiv u^{\mathrm{bulk}}_K/\epsilon_{\mathrm{vac}}}, which defines the ratio of bulk kinetic energy over the vacuum energy. The third parameter is the bubble wall velocity $v_b$. It is shown that the efficiency parameter $\kappa$ depends on the bubble wall velocity $v_b$ and $\alpha_N$~\citep{Steinhardt1982,Kamionkowski:1993fg,Espinosa:2010hh}.
%
\begin{comment}
\begin{equation}
 \alpha_N\equiv\frac{\epsilon_{\mathrm{vac}}}{\rho_{\mathrm{rad}}}
  \qquad\textrm{and}\qquad
  \kappa\equiv\frac{u^{\mathrm{bulk}}_K}{\epsilon_{\mathrm{vac}}}\,.
\end{equation}
%
\end{comment}
%
%
With equipartition between magnetic and kinetic energy, the fraction of
magnetic energy to the radiation energy
\m{f_E=u_K/\rho} becomes \m{f_E=\kappa\alpha_N}, hence
\m{(B^{\mathrm{max}}_{\lambda,*}/B_{I,*})^{2}=1/2\kappa\alpha_N} in Eq.~(\ref{eq:fmin}). 
%
\begin{comment}
gives
\begin{equation}
 f_*\geq f^{\mathrm{min}}_{*}
  =\left(5.6\times10^{-11}\right)\frac{\lambda_\mathrm{EW}}{\lambda_{I,*}}
  \frac{1}{2\kappa\alpha_N}\,.
\end{equation}
\end{comment}
%
Following the work of \citep{Espinosa:2010hh} we can explore the parameter space for 
$f^{\mathrm{min}}_{*}$, independently of a specific particle physics model of the phase transition.
The results are shown in Fig.~\ref{fig:PT_Plot}.
For example, for a weak phase transition \m{\alpha_N=0.01} and subsonic bubble wall velocity \m{v_b=0.1},  we find the minimum helicity fraction \m{f^{\mathrm{min}}_{*}\sim10^{-3}} at magnetogenesis. Whereas for a strong phase transition \m{\alpha_N\approx1} and supersonic bubble wall velocity \m{v_b=0.9}, we find the minimum helicity fraction \m{f^{\mathrm{min}}_{*}\sim10^{-8}} at magnetogenesis i.e. much larger than the SM predictions from electroweak baryogenesis, where \m{\lambda_{I,*}/\lambda_{\mathrm{EW}}\sim10^{-2}} for the EWPT was used in both cases.

\begin{comment}
For a given wall velocity there is a maximum value for $\alpha_N$, \m{\alpha_N^{\mathrm{max}}\simeq\frac13(1-v_b)^{-13/10}} (Espinosa et al 2010), hence this will also give us a maximum $\kappa$ and therefore the lowest bound on fraction $f_E^{\mathrm{min}}$.
\end{comment}

%\begin{comment}
%
%
%\end{comment}

\section{Conclusions}
%\paragraph{Conclusions:}

First-order phase transitions can generate magnetic fields in the early Universe.  
Under early Universe conditions with very small chemical potentials the QCDPT is a smooth transition \citep{Aoki:2006we} whereas the EWPT could be first-order in certain Standard Model (SM) extensions ~\citep{Laine:1998qk}. Inflationary magnetogenesis \citep{Turner:PMF}, which is also beyond the SM, is another popular mechanism to explain void magnetic fields. Hence, the apparent observations of void fields from $\gamma$-ray observations seem to be a signature of physics beyond the SM or of new mechanisms which excite magnetic helicity \citep{Boyarsky:2011uy,Boyarsky:2015faa}. If the constraints on void fields prove to be conclusive, then it is likely that magnetic helicity must play an important role. Here we show that magnetic fields generated at the EWPT must have significantly more helicity than that produced by electroweak baryogenesis in order to explain the extragalactic magnetic fields. To reach this conclusion we have assumed the magnetic decay laws of \citep{Banerjee:2004df,Campanelli:2007tc,Campanelli:2013iaa} and considered the simplest magnetic helicity estimates of \citep{Vachaspati:2001nb}.
Our assumptions on the decay rates could be challenged due to new results of \citep{Kahniashvili:2012uj,Brandenburg:2014mwa}.

%\begin{comment}
%\begin{acknowledgments}
\section*{Acknowledgments}
We thank G. Sigl and T. Konstandin for helpful discussions and comments. This work was supported by the Deutsche Forschungsgemeinschaft (DFG) through the collaborative research centre SFB 676 Particles, Strings, and the Early Universe.
%\end{acknowledgments}
%\end{comment}

%%---------------------------------------------------------
%%---------------------------------------------------------

\bsp
\label{lastpage}

%\section*{References}
\bibliographystyle{mn2e}
\bibliography{references_Mag}

\end{document}